\documentclass{article}
\usepackage{arxiv}
\usepackage[utf8]{inputenc} % allow utf-8 input
\usepackage[T1]{fontenc}    % use 8-bit T1 fonts
\usepackage{setspace}
\usepackage{hyperref}       % hyperlinks
\usepackage{url}            % simple URL typesetting
\usepackage{booktabs}       % professional-quality tables
\usepackage{amsfonts}       % blackboard math symbols
\usepackage{nicefrac}       % compact symbols for 1/2, etc.
\usepackage{microtype}      % microtypography
\usepackage{lipsum}		% Can be removed after putting your text content
\usepackage{graphicx}
\usepackage{natbib}
\usepackage{lineno}
\usepackage{dsfont}
\usepackage{doi}
\usepackage{amsthm,amsmath,amssymb}
\usepackage{mathtools}
\usepackage{graphicx,bm}
\usepackage{color}
\usepackage{lscape}
\usepackage{rotating}
\usepackage{setspace}
\usepackage{algorithm}
\usepackage{algpseudocode}
\usepackage{epigraph}

\title{Smooth Reduced Rank Regression with P-splines}

\author{ \href{https://orcid.org/0000-0001-7308-6210}{\includegraphics[scale=0.06]{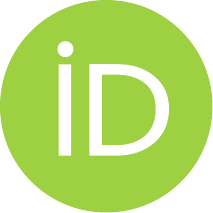}\hspace{1mm}Mark de Rooij}\\
 	Methodology and Statistics department, Leiden University\\
  	The Netherlands \\
  	\texttt{rooijm@fsw.leidenuniv.nl} \\
   }

\renewcommand{\headeright}{}
\renewcommand{\undertitle}{}
\renewcommand{\shorttitle}{Smooth Reduced Rank Regression}

\makeatletter
\input{size12.clo}
\makeatother

\begin{document}
\maketitle
%\linenumbers

\begin{abstract}
Linear regression is one of the core statistical tools used for analysis of data. In the era of statistical learning, linear regression has been expanded into two directions. The first is \emph{regularization}, where penalties are added to the loss function to obtain more stable or sparse solutions. This direction is especially useful for high dimensional data sets, where a researcher has many predictors. The second direction is \emph{basis expansion}, such as with spline or kernel functions, where the linearity assumption is dropped. In practice, empirical researchers often collect multiple outcome variables. Regression models, either linear, regularized, or expanded, can be fitted to each of these outcome variables, but such an approach does not take into account the associations among the response variables. Reduced rank regression is a multivariate regression tool that takes into account and models the association among the response variables. In this paper, we develop and test a B-spline basis expansion for reduced rank regression, a regression model for multiple outcome variables where we penalize the coefficients to obtain a smooth fit, as in P-splines. This approach is useful for the analysis of data sets with multiple outcomes and a relatively small number of predictors. An block-relaxation algorithm is developed for parameter estimation and we discuss ways for penalty parameter tuning. We develop visualization tools for model interpretation based on biplot methodology for ``\emph{all predictor - all response}'' relationships and partial dependence plots for interpreting ``\emph{single predictor - single response}'' relationships. With several experimental gauges we test the algorithm and show it works adequately. Afterwards, we analyze three empirical data sets: The first example highlights the importance of penalty parameter tuning and shows that the BIC performs better than the AIC in terms of smoothing; The second example shows in detail the triplot and partial dependence plots and their relationships; The third shows an application to ecological momentary assessment data, a type of data that becomes more and more important in psychological research. We conclude this paper with a discussion. 
\end{abstract}


\newpage

\section{Introduction}

% In this paper, we will describe a general methodology for the case where we are interested in the analysis of dependence, that is, we like to see the dependence of a set of variables on another set of variables, as in regression methods. Such an analysis of dependence can be contrasted to an analysis of interdependence \citep{gifi1990nonlinear}, the latter treats the variables in a symmetric way, as in correlation and association analysis. 

Linear regression is one of the general workhorses of statistics. In linear regression, a numeric response variable is linked to a set of predictor variables, where each predictor variable is given a weight such that the estimated response variable is as close as possible to the observed response variable in a least squares sense. In the last decades the linear regression model has been adapted in two directions. For the case that there are many predictor variables, regularization techniques, such as ridge regression, lasso regression, and elastic net regression, have been proposed, where a penalty is added on the size of the estimated weights. The other direction, is to relax the linearity assumption and to allow for nonlinear relationships by basis expansions. Examples of basis expansions are polynomial transformations of the original predictors, kernel functions, or spline representations. These two directions are discussed in several statistical learning textbooks such as \cite{friedman2001elements}, \cite{james2013introduction}, or \cite{berk2020statistical}. In this paper, we will focus on basis expansion with spline functions.  

% @book{berk2020statistical,
%   title={Statistical learning from a regression perspective (3rd edition)},
%   author={Berk, Richard A.},
%   year={2020},
%   publisher={Springer}
% }

Usually, researchers have multiple response variables and multiple predictor variables. Typically, researchers analyze the data in a univariate way, one response variable at a time. Such separate analyses, however, ignore the dependencies among the response variables. \cite{fish1988multivariate} argued that it is important to analyze the data using multivariate methods because most outcomes have multiple causes and most causes have multiple effects. With $P$ predictor and $R$ continuous response variables, we could fit a multivariate regression model
\[
\bm{Y} = \bm{1m}' + \bm{XA} + \bm{E},
\]
that relates the set of response variables with the set of predictor variables. The estimated coefficients of such a regression model, when estimated with least squares or maximum likelihood techniques, are equal to the coefficients from $R$ separate regression models, one for each response variable. Hence, the fact that the response variables are likely to be related does not play a role in the estimation as no information about the associations is taken into account \citep[][chapter 1]{reinsel2022multivariate}. A truly multivariate model would take such information into account and model in some way the associations among the response variables. 

Various approaches can be considered to develop a truly multivariate model that 1) takes into account and models the associations among the response variables, 2) reduces the number of parameters, and 3) facilitates the interpretation of the final model. The reduced rank regression (RRR or R$^3$) model also known as redundancy analysis is such an approach. This model was first introduced by \cite{anderson1951estimating} and further developed by several authors \citep{izenman1975reduced, tso1981reduced, davies1982procedures, wollenberg1977redundancy}. An overview of developments for reduced rank models can be found in \cite{reinsel2022multivariate}. The key idea is to write the matrix of regression coefficients ($\bm{A}$) as a product of two matrices of lower dimensionality, namely the regression weights ($\bm{B}$) and the factor loadings ($\bm{V}$), that is,
\[
\bm{A} = \bm{BV}',
\]
where $\bm{B}$ is a $P \times S$ matrix and $\bm{V}$ is an $R \times S$ matrix. The resulting matrix $\bm{A}$ has rank $S$ whereas the matrix in the multivariate regression model has rank $\min(P, R)$, hence the name reduced rank regression. We will use the names rank or dimensionality interchangeable. The user has to choose the required rank or a model selection procedure for choosing an optimal $S$ should be employed. When $S = \min(P, R)$  the reduced rank regression model becomes equal to the multivariate regression model. When $S < \min(P, R)$ the number of parameters is reduced.

The low rank structure implies that the response variables are dependent on the predictor variables through a small number of latent variables. These latent variables are defined as $\bm{U} = \bm{XB}$, that is, they are linear combinations of the predictor variables. These $S$ latent variables are shared among the responses and as such the reduced rank regression model with reduced $S$ is a truly multivariate model taking into account the associations among the response variables \citep{luo2018leveraging}. When $S < \min(P, R)$ the model implies associations among the responses. To see that, take $S = 1$. In that case the model implies associations among \emph{all} response variables: positive associations when the loadings of two response variables have the same sign, negative when the sign differs. When $S = 2$, the model might break up the response variables in two groups, one group with zero loadings on the second dimension, the other group with zero loadings for the first dimension. Variables that pertain to a certain dimension have model implied associations (positive if the sign is equal, negative otherwise). The model implied association for pairs of variables pertaining strictly to different dimensions is null. In data analysis, we usually do not find estimates exactly equal to zero. When $S > 2$, the model becomes more and more flexible in its ability to represent associations. Ultimately, when $S = R$ each response variable might pertain to a certain dimension and the model does not imply the responses to be correlated.

% Reduced rank regression can also be considered a constraint principal component analysis \citep{takane2013constrained}.  In principal component analysis the centered data matrix $\bm{Y}_c = \bm{Y} - \bm{1m}'$ is usually decomposed in a set of object scores and a set of variable loadings, that is,
% \[
% \bm{Y}_c = \bm{UV}' + \bm{E}.
% \]
% Here both the matrices $\bm{U}$ and $\bm{V}$ are assumed to have a few (i.e., $S$) columns. 
% In reduced rank regression, the object scores $\bm{U}$ of principal component analysis are constrained to be a linear combination of the predictor variables $\bm{U} = \bm{XB}$.

% The reduced rank regression model has been generalized in several directions. First, the reduced rank model has been generalized for non-numeric response variables. \cite{yee2015} extended the models for response variables in the exponential family, whereas \cite{derooij2023new} developed an extension specifically for binary response variables. For ordinal variables, \cite{derooijbreemerwoestenburgbusing2022}, recently described a reduced rank model with a cumulative link function. Both \cite{luo2018leveraging} and \cite{derooij2025gmr3} proposed reduced rank regression models for mixed outcomes. Second, various authors proposed to regularize the reduced rank regression model, see for example \cite{chen2012sparse, chen2016sparse, kobak2021sparse}, and chapter 13 of \cite{reinsel2022multivariate} for sparse approaches and \cite{} and chapter 10 of \cite{reinsel2022multivariate} for reduced rank regression with a nuclear norm penalty for automatic dimension selection.

In reduced rank regression there is a linearity assumption. The predictor variables have a linear effect on the outcomes. The linearity assumption is a strong assumption and often not tenable. An relatively easy extension, is to include polynomial terms in the design matrix $\bm{X}$. Such polynomial terms can be included for each of the predictor variables. The result is a large model selection problem, as for each predictor variable an optimal degree for the polynomial has to be found. Including polynomial terms in the design matrix forms an example of basis expansion. Other forms of basis expansion are based on step functions, kernel functions, piecewise polynomials, the truncated power basis, or other spline bases \citep[see, for example, ][]{james2013introduction}. All such basis expansions lead to model selection problems. 

% \citep{eilers1996flexible, eilers2021practical}. P-splines are defined by a large B-spline basis (i.e., a basis with many knots) and a regularization penalty on the difference of estimated coefficients. With P-splines the model selection problem is reduced to tuning the penalty parameter. 

% We agree with the first sentence of the book \citep{eilers2021practical}, stating "\emph{This is a book about P-splines, our favorite smoother, and the best one you can find}".  

In this paper, we will extend the reduced rank regression model with P-splines. \cite{eilers1996flexible} first proposed P-splines, the combination of a large B-spline basis with penalized estimation of the coefficients to obtain to smooth fitted curves by choosing the penalty parameter wisely. In this approach, the model selection problem is reduced to tuning a penalty parameter. \cite{eilers1996flexible} show that this tuning can either be performed using cross-validation or by employing an information criterion. \cite{eilers2021practical} provide a concise overview of P-splines and discussed many applications and generalizations. In this paper, our goal is to develop a smooth reduced rank regression approach that largely follows the P-spline approach to (generalized) additive models, as discussed in Section 4.1 of \cite{eilers2021practical}, but then for multivariate outcomes with a reduced rank restriction on the coefficient matrix. We will propose this model, describe an algorithm for parameter estimation, discuss model selection and introduce graphical tools for interpreting the estimated model.  

In the next section, we lay out our new methodology. We introduce the B-spline basis, the penalties, an algorithm for parameter estimation, discuss model selection, and some implementation choices. In Section \ref{sec:visualisation}, we show two types of visualizations of our modeling results. The first type of visualizations is based on biplot methodology \citep{gower1996biplots, gower2011understanding} resulting in a single graphical representation of all predictors and all responses. The second type of visualizations are partial dependence plots, that are graphs per single predictor and single response combination. In Section \ref{sec:gauges}, we show some experimental simulated data sets to which we apply our methodology. The gauges have highly non-linear trajectories as well as strict linear trajectories and combinations thereof. We show that our methodology deals adequately with these limiting situations. In Section \ref{sec:examples}, we show three empirical examples. We conclude this paper with some discussion. 

\section{RRR with P-splines}

In this Section, we lay out the general proposed methodology. We start with a $N \times R$ matrix $\bm{Y}$ of responses and another $N \times P$ matrix $\bm{X}$ of predictors. We assume both have numeric variables. Furthermore, we assume the responses are centered and we will denote the means by $\bm{m}_y$, estimated as $\hat{\bm{m}}_y = (\bm{1}'\bm{1})^{-1}\bm{1}'\bm{Y}$. To make this centering assumption explicit, we define 
\[
\bm{Y}_c = \bm{Y} - \bm{1}\hat{\bm{m}}'_y, 
\]
and use $\bm{Y}_c$ in our formulae. 

We start in the next section with an explanation of the B-spline basis functions and define the smooth reduced rank regression model, then discuss penalties on differences of coefficients,  and some implementation choices we made, describe an algorithm, and discuss model selection. 

\subsection{B-spline basis} 

\begin{figure}
    \centering
    \includegraphics[width = 0.8\textwidth]{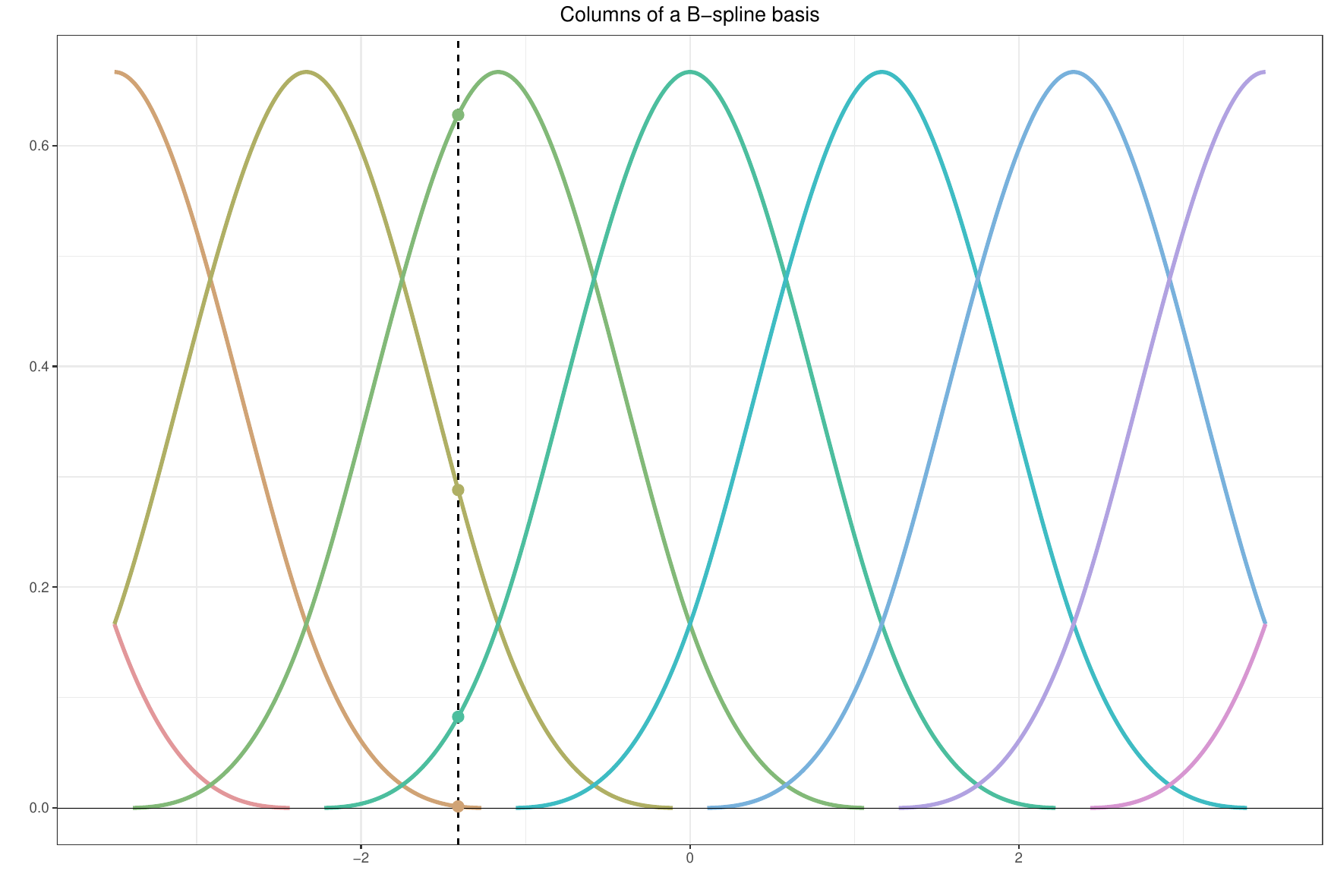}
    \caption{B-spline basis functions for a variable with range -3.5 to 3.5 with degree 3 and 7 knots.}
    \label{fig:bsplines}
\end{figure}

For flexible fitting of a regression line often a basis expansion is used, where a variable is transformed from $\mathbb{R}$ to $\mathbb{R}^p$. One such basis expansion is given by B-splines \citep{deboor1978practical}. We illustrate this expansion in Figure \ref{fig:bsplines}, 
% a figure in a similar way as in \citep[][Figure 2.3]{eilers2021practical}, 
for a variable whose minimum is minus 3.5 and maximum plus 3.5. A spline basis is defined by two characteristic, its \emph{degree} and it's \emph{number of segments} or, equivalently, the \emph{number of knots}. The number of knots minus one equals the number of segments. The higher the number of segments (or knots) the higher the flexibility of the smooth curve. The higher the degree of the B-splines, the smoother the curve. 

Each B-spline consists of polynomial segments that begin and end at specific values of the variable, called knots. At the knots two of these segments meet, such that the value of the B-spline, but also all the derivatives upto the order `degree minus one' are equal. So, for a cubic spline the polynomial segments each have degree 3 and the values, first, and second derivatives are equal at the knots. The number of polynomial segments for each B-spline equals the degree plus one.

In Figure \ref{fig:bsplines} a cubic spline is defined, that is a spline basis with degree three. The number of segments equals six (equivalently, there are 7 knots). Therefore, the B-spline basis matrix $\bm{Z}$ has nine columns (degree + number of segments). Each row corresponds with the values of the nine B-splines corresponding to a single original value of the predictor. With a degree 3 spline, only four of these values in a row are non-zero. For the predictor value $x= -1.41$, indicated with the vertical dotted line in Figure \ref{fig:bsplines}, it is shown that the values on four B-splines are non-zero (i.e., represented by the dots on the B-splines). The values for all other B-splines at this value of the predictor equal zero. The spline basis matrix $\bm{Z}$ is therefore a sparse matrix. 
%The sum of the four values, for any value of the variable, equals one. As the other values are zero, the row sums of $\bm{Z}$ are one as well.

With $P$ predictor variables, we need a basis expansion for each predictor, that is, for predictor variable $p$, we transform $\bm{x}_p$ to $\bm{Z}_p$. As a result, we have $P$ of these matrices, and we simply concatenate them horizontally to obtain the matrix $\bm{Z}$, that is
\[
\bm{Z} = \left[\bm{Z}_1, \ldots, \bm{Z}_p \right].
\]
Now, we may write our smooth reduced rank regression model as
\[
\bm{Y}_c = \bm{Z}\bm{BV}' + \bm{E},
\]
where the matrix $\bm{B}$ is defined as $\bm{B} = \left[ \bm{B}'_1, \ldots, \bm{B}'_P \right]'$, with $\bm{B}_p$ the coefficients pertaining to the $p$-th predictor variable and with $\bm{E}$ a matrix with residuals $e_{ir}$. 

\subsection{Penalties} 

The smoothness of the fitted curve does not only depend on the degree of the spline basis but also on the estimated coefficients \citep[][Figure 2.8]{eilers2021practical}. Even with a large number of segments, but a very regular estimated set of coefficients, the fitted curve becomes smooth. \cite{eilers1996flexible} used this property to define their P-spline approach. 

To regulate the smoothness, typically a large number of segments is chosen and the smoothness of the fitted curve is managed by penalizing the differences between adjacent coefficients. These differences might be of the order one, two, or higher and can be defined by a differencing matrix $\bm{D}$. The differencing matrix of order one is defined as
\[
\bm{D}_1 = \left[
\begin{array}{rrrrr}
-1 & 1 & 0 & 0 & \cdots \\
0 & -1 & 1 & 0 & \cdots \\
0 & 0 & -1 & 1 & \cdots \\
\vdots & \vdots & \vdots & \vdots & \ddots
\end{array} \right],
\]
such that $\bm{D}_1\bm{B}$ gives the differences in coefficients of two consecutive rows. Similarly, the differencing matrix of order two is defined as
\[
\bm{D}_2 = \left[
\begin{array}{rrrrrr}
1 & -2 & 1 & 0 & 0 &\cdots \\
0 & 1 & -2 & 1 & 0 & \cdots \\
0 & 0 & 1 & -2 & 1 & \cdots \\
\vdots & \vdots & \vdots & \vdots & \vdots & \ddots
\end{array} \right].
\]
In P-splines these differences are penalized to make them as small as possible. Smaller differences result in smoother curves. Therefore a penalty parameter is introduced that gives a weight to these differences where larger value indicate more weight. In a model with $P$ predictor variables, a differencing matrix and a corresponding penalty parameter is needed for each predictor. 

% It is useful to know the limiting behavior as the penalty parameter becomes very large. For univariate outcomes, \cite{eilers2021practical} establish that when the order equals one, the fitted curve will become a horizontal line as the penalty parameter increases; when the order equals two, the fitted curve will become linear as the penalty parameter increases; when the order equals three, the fitted curve will become quadratic as the penalty parameter increases; when the order equals $d$, the fitted curve will become a polynomial of degree $d - 1$ as the penalty parameter increases. 

\subsection{Choices in our implementation}

In the next Section, we will describe an algorithm. We implemented our Algorithm in R. The functions are available from the github page of the author. In the implementation, we made several simplifying choices. We use the same number of knots (default equals 21) and the same degree (default is 3) for every predictor variable. We use the same differencing matrix (default is second order differences) for all predictor variables. The sequence of penalty parameters is typical linear on the $\log_{10}$ scale. % starting from, say, -2 and till 5. 
We also use these simplifications in our description below. The methodology we describe is more general, but the coding would become rather complicated in some cases. 

\subsection{Algorithm} 

The parameters of our smooth reduced rank regression model are the two matrices $\bm{B}$ and $\bm{V}$. We employ a penalized least squares loss function composed of the sum of the squared residuals ($e_{ir}$) plus the sum of squared differences of the coefficient matrix is minimized. 

We will develop a block relaxation algorithm for estimating the two sets of parameters $\bm{B}$ and $\bm{V}$ where we alternate between updating $\bm{B}$ and $\bm{V}$ keeping the other set fixed. Such an algorithm monotonically converges to a minimum. 

The loss function is composed of the least squares part and a penalty part. The penalty parameter for predictor variable $p$ is indicated by $\lambda_p$. With $P$ predictor variables we therefore have the vector of penalty parameters $\bm{\lambda} = [\lambda_1, \ldots, \lambda_P]$ and we define the diagonal matrix $\bm{\Lambda} = \mathrm{diag}(\bm{\lambda})$. For the penalty part, we define the matrix
\[
\bm{P}_{\lambda} = \bm{\Lambda} \otimes \bm{D}'\bm{D} + \kappa \bm{I}
\]
with $\kappa$ a small constant, say $\kappa = 10^{-5}$, to keep the sum of squared coefficients as small as possible for identification. With this penalty part, the loss function we need to minimize is 
\begin{eqnarray*}
\mathcal{L}(\bm{B}, \bm{V} | \lambda ) = \| \bm{Y}_c - \bm{Z}\bm{B}\bm{V}' \|^2 
+ \mathrm{tr} \bm{B}'\bm{P}_{\lambda}\bm{B}.
\end{eqnarray*}
The parameters are not identified because we can find a new set of parameters with the same fit as $\bm{B}\bm{V}' = \bm{BT}\bm{T}^{-1}\bm{V}' = \bm{B}^*(\bm{V}^*)'$ for any non-singular matrix $\bm{T}$. To identify the solution we require $\bm{V}'\bm{V} = \bm{I}$. 

As alluded, we alternate between updating the two sets of parameters. We first consider $\bm{V}$ fixed, so $\mathcal{L}$ is only a function of $\bm{B}$. Minimizing this function is relatively straightforward. Using results form \cite{penrose1956best} as described in \cite{tenberge1993}, the update for $\bm{B}$ is given by
\[
\bm{B}^+ = (\bm{Z}'\bm{Z} + \bm{P}_{\lambda})^{-1}\bm{Z}'\bm{Y}_c\bm{V}. 
\]
Note that $(\bm{Z}'\bm{Z} + \bm{P}_{\lambda})$ does not change during the iterations so we only have to compute the inverse once.  
   
When updating $\bm{V}$, we consider $\bm{B}$ fixed. The penalty part of the loss function in this case can be considered a constant, so that we may write the loss function as
\begin{eqnarray*}
\| \bm{Y}_c - \bm{UV}' \|^2,
\end{eqnarray*}
where $\bm{U} = \bm{Z}\bm{B}$. This function needs to be minimized subject to the imposed identification constraint $\bm{V}'\bm{V} = \bm{I}$. The update can be found using Kristof's upper bound as has been shown by \cite{tenberge1993}. Therefore, we first take the singular value decomposition of the matrix
\[
\bm{U}'\bm{Y}_c = \bm{P}\bm{\Phi}\bm{Q}'
\]
from which we compute the update
\[
\bm{V}^+ = \bm{Q}_S\bm{P}'_S,
\]
where $\bm{P}_S$ and $\bm{Q}_S$ are the left and right singular vectors corresponding to the $S$ largest singular values.

The algorithm alternates between updating of $\bm{B}$ and $\bm{V}$, in each iteration the loss function decreases. As the loss function is convex, this alternating algorithm converges to the global minimum of the loss function. We stop iterating when the change in the loss function is smaller than a predetermined convergence criterion, which is usually within a few iterations. 

\subsection{Model selection}

When analyzing a data set, a researcher has to choose two parameters: The dimensionality $S$ and the penalty values $\lambda_p$. The latter is a vector of length equal to the number of predictor variables. 

We propose to select values both the dimensionality as well as the penalty values based on information criteria, the AIC and the BIC. Both information criteria trade-off model fit with model complexity. The definitions are
\[
\mathrm{AIC} = \mathcal{D} + 2 \mathrm{ED}
\]
and 
\[
\mathrm{BIC} = \mathcal{D} + \log(N) \mathrm{ED}.
\]
Both definitions include $\mathcal{D}$, the deviance, and ED, the effective model dimension. For our model $\mathcal{D}$ is defined as
\[
\mathcal{D} = \frac{\sum_i \sum_r  (y_{ir} - \mu_{ir})^2}{\sigma^2} + NR \log\sqrt{2\pi \sigma^2}
\]
where $\mu_{ir}$ are the estimated expected values and $\sigma^2$ is the variance of the residuals estimated as
\[
\hat{\sigma}^2 = \frac{1}{N R}
\sum_{i = 1}^N \sum_{r = 1}^R e^2_{ir}.
\]
ED represents the effective model dimension, defined as
\[
\mathrm{ED}_{\lambda} = \mathrm{tr} \left[ (\bm{Z}'\bm{Z} +\bm{P}_{\lambda} )^{-1} \bm{Z}'\bm{Z} \right] + (R - S)S, 
\]
consisting of two parts. The latter part $(R - S)S$ concerns the number of estimated parameters in $\bm{V}$ minus the number of identification constraints. The first part, concerns the number of effective parameters in estimating $\bm{B}$ with the smoothness penalty. This terms generalizes a similar term in \cite{eilers2021practical}. 

For several values of $S$ we find the optimal penalty parameter values (i.e., the $\lambda$'s) using either the AIC or BIC. Afterwards, we compare the values for the different values of $S$ and again choose the one that minimizes the information criterion. 

\section{Graphical Representations}\label{sec:visualisation}

\subsection{Triplots for all-2-all relationships}

When $S = 2$ the modeling results can be represented in a triplot, a graphical representation that shows the predictor variables, the response variables, and the observations in one display. When $S > 2$. we can use the same methodology but for pairs of dimensions, say 1 and 2, or 1 and 3. 

We built our triplots on earlier derived results about biplots as described in \cite{gabriel1971biplot, gower1996biplots, gower2011understanding}. Biplots are graphical representations of data that show two pieces of information, the (response) variables and observations, in a plot with properties similar to standard scatterplots. These biplots are based on principal component analysis (PCA) of a data matrix ($\bm{Y}_c$). For our purposes it is important to understand that reduced rank regression is a constrained PCA, where the object scores of PCA are functions of the predictor variables.
\cite{braak1994biplots} generalized these biplots to triplots for reduced rank regression. Triplots represent, besides the observations and the response variables, also the predictor variables. Triplots therefore represent three pieces of information (hence the tri). In \cite{braak1994biplots} the predictor variables are represented by straight lines as the triplot represents the linearity assumed in standard reduced rank regression. We will represent the predictor variables with smooth trajectories in the triplot. 

In our elaboration, we first introduce a graphical representation of the predictor variables and the observations where we discuss the process of interpolation. Thereafter, we show a graphical representation of the observations and response variables for which we discuss the process of prediction. Usually, these two graphical representations are combined in one triplot such that we can arrive at conclusions about the relationships between the predictor variables and the response variables. Such triplots will be shown in later sections. For simplicity, we focus on a two-dimensional solution for a data set with two predictor variables (X1 and X2) and three response variables (A, B, C).  

\begin{figure}
    \centering
    \includegraphics[width = 1.0\textwidth]{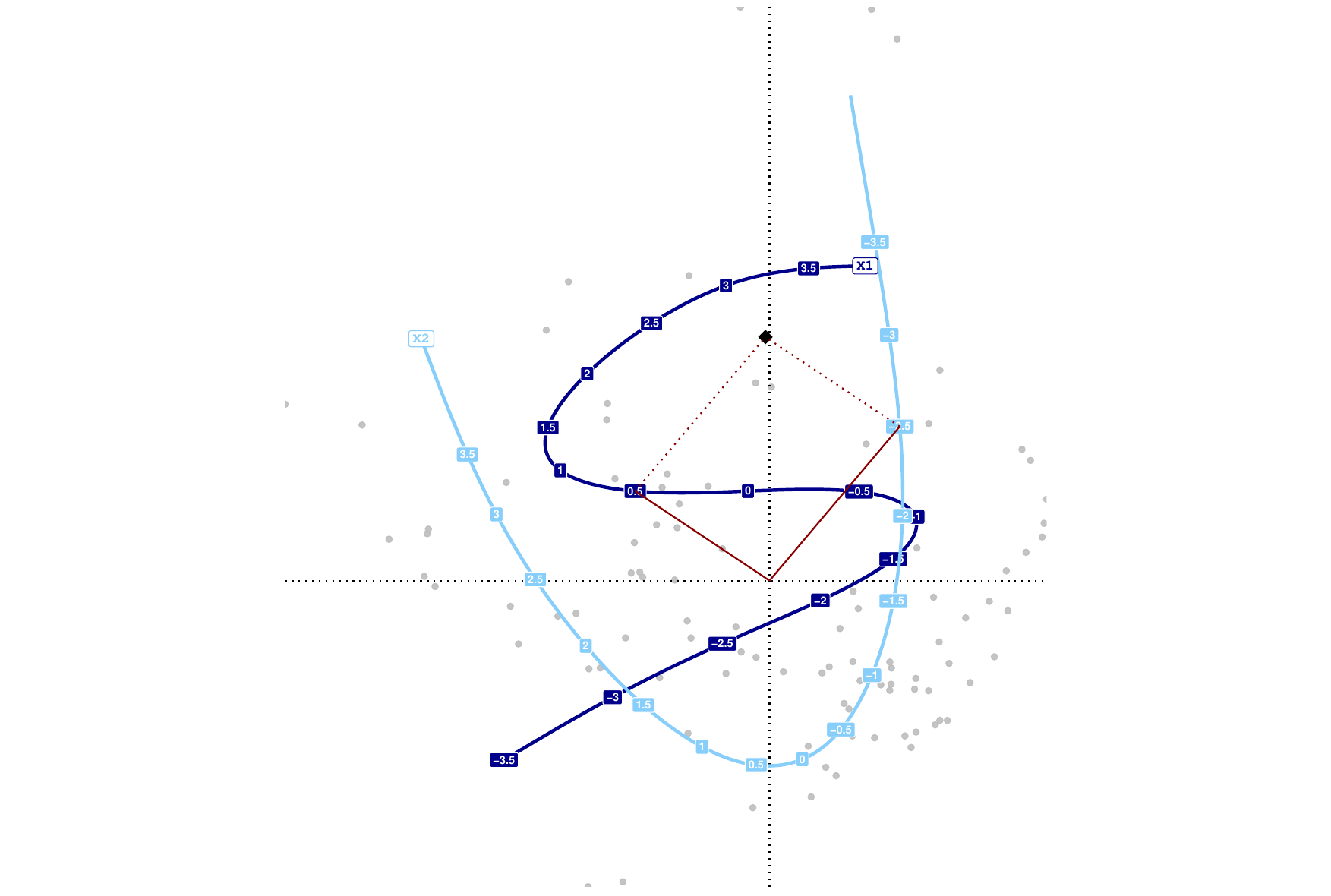}
    \caption{Explanation of triplot I: Observations and prediction variable trajectories.}
    \label{fig:biplot1}
\end{figure}

Our algorithm estimates the weights ($\bm{B}$) and the loadings ($\bm{V}$). Remember, we centered the response variables before analysis, so we also have the means of these ($\bm{m}$). With the estimated weights, $\hat{\bm{B}}$ we can obtain the positions of the observations $\hat{\bm{U}} = \bm{Z}\hat{\bm{B}}$. Each row in this matrix represents the coordinates for an observation. In Figure \ref{fig:biplot1}, the observations are shown by the grey dots. 

In this plot the horizontal and vertical axis are shown by dotted lines. These are added for explanatory purposes only and usually not drawn.  

Also shown in Figure \ref{fig:biplot1} are the variable trajectories for the two predictor variables. These are drawn as follows. We first define for each predictor variable $p$ an auxiliary vector $\tilde{\bm{x}}_p$ with a long sequence of values from the minimum to the maximum observed value. Using the same basis expansion as in the algorithm, $\tilde{\bm{x}}_p$ is transformed to $\tilde{\bm{Z}}_p$. With this auxiliary matrix and the estimated weights we can compute 
$\tilde{\bm{Z}}_p\hat{\bm{B}}_p$ that forms a smooth sequence of points in the two-dimensional space that forms a trajectory for predictor $p$. In Figure \ref{fig:biplot1} we see two such trajectories, one for each predictor variable. The label of the predictor variable is printed at the maximum value. So, X1 has an S-shaped trajectory where low values are represented in the bottom left of the representation and the trajectory runs like a mountain road to the upper right corner of the visualization. For X2 the trajectory follows a U-shaped pattern, starting in the upper right corner of the visualization and ending in the upper left corner. 

We add valued markers to the trajectories. Therefore, a series of rounded numbers within the range of the variable is defined. For X1 these numbers run from -3.5 to 3.5 in steps of a half. The rows in $\tilde{\bm{Z}}_1$ corresponding to these values are selected. Multiplying this selection with the estimated weight provides the coordinates of the points where we add the valued markers.  

The smooth reduced rank regression model is an additive model. Therefore, we are able to obtain the position of a person from the two variable trajectories. Suppose this person has scores 0.5 on X1 and -2.5 on X2. Vectors can be drawn from the origin (i.e., the position where the horizontal and vertical axes cross) to the valued markers for these two variables. These are shown by solid red lines in the display. Because we defined an additive model, we simply have to add these two vectors to obtain the position for the person in the visualization. The addition is shown by the red dotted lines and the person is indicated by the larger black diamond symbol. The coordinates of this person with scores 0.5 and -2.5 on the two predictor variables is $\bm{u} = [-0.03, 1.73]$. 

\begin{figure}
    \centering
    \includegraphics[width = 1.0\textwidth]{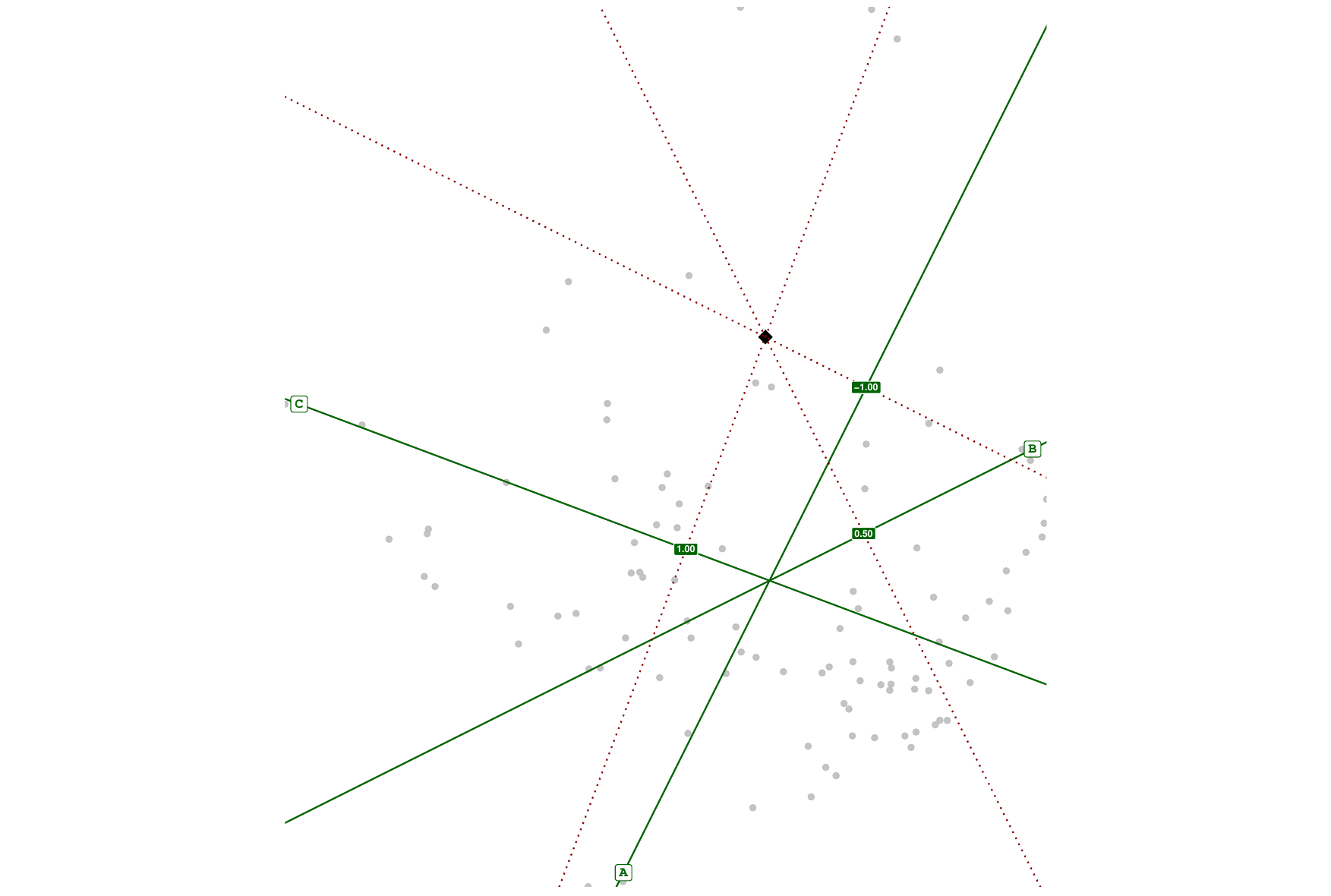}
    \caption{Explanation of triplot II: Observations and response variable axes.}
    \label{fig:biplot2}
\end{figure}

Now, let us go too the second part of the explanation of the triplots, for which we use Figure \ref{fig:biplot2}. This visualization shows the same grey dots for the observations as well as three variable axes for the response variables (A, B, C). The estimated loadings in this case are
\[
\hat{\bm{V}} = \left[ \begin{array}{rr}
\frac{1}{3} & -\frac{2}{3} \\
1 & \frac{1}{2} \\
 -\frac{2}{3}  &  \frac{1}{4} \end{array} \right].
\]

The variable axes cross each other in the origin of the visualization and the direction of the axes for variable A is $\hat{v}_{A2}/\hat{v}_{A1}$. So, for variable A, the first row of the estimated loading matrix, the direction is -2. The direction for the other two response variables can be obtained from the other rows of the loading matrix. We add a name marker at the positive end of the variable axes.

We saw that the observation with observed scores 0.5 and -2.5 is presented by a point with coordinates 
$[-0.03, 1.73]$. To obtain the predicted scores on the response variables we need to project the point onto the response variable axes. This corresponds with computing the inner product $\bm{u}'\bm{v}_r$, the values are
\begin{itemize}
\item Variable $A$: 
\[
\bm{u}'\bm{v}_A = \left[ -0.03, 1.73 \right] \left[ \begin{array}{r} -0.33\\ -0.66 \end{array}\right] = -1.16
\]
\item Variable $B$:
\[
\bm{u}'\bm{v}_B = \left[ -0.03, 1.73 \right] \left[ \begin{array}{r} 1.00 \\ 0.50 \end{array}\right] = 0.88
\]
\item Variable $C$:
\[
\bm{u}'\bm{v}_C = \left[-0.03, 1.73 \right] \left[ \begin{array}{r} -0.66 \\ 0.25 \end{array}\right] = 0.42
\]
\end{itemize}
above the average for the three variables. Suppose the averages are 0.16, -0.38, and 0.58, for variables A, B and C, respectively then the predicted values are -1, 0.5, and 1. We can add the valued markers -1, 0.5, and 1 to the three points of projection. More generally, we can add in a similar way valued markers to the response variables for pretty numbers in the range of the response variables \citep[][Chapter 2, section 2.3]{gower2011understanding} which allows us to read of the predicted values for any position in the visualization. We show these in the upcoming sections. 

\subsection{Partial dependency plots for one-2-one relationships}

Apart from the biplots, we can also visualize the estimated relationship between predictor $p$ and response variable $r$. These visualizations are similar to the visualizations for penalized (generalized) additive models as shown in Section 4.1 of \cite{eilers2021practical}. On the horizontal axes, the predictor variable is shown, on the vertical axes the partial response. The data together with the smooth fitted curve are displayed in such a plot. 

For the $r$-th response variable, we can write our model as
\[
\bm{y}_r = \bm{Z}\bm{Bv}_r + \bm{e}_r = \bm{Z}_p\bm{B}_p\bm{v}_r + \bm{Z}_{-p}\bm{B}_{-p}\bm{v}_r + \bm{e}_r,
\]
where $\bm{Z}_p$ is the part of $\bm{Z}$ corresponding to the $p$-th predictor 
and $\bm{Z}_{-p}$ the remainder of the matrix. A similar breakdown is created for the matrix $\bm{B}$. Let us define the partial residuals as
\[
\bm{z}_{r, -p} = \bm{y}_r - \bm{Z}_{-p}\bm{B}_{-p}\bm{v}_r.
\]
This allows us to make a scatterplot of $\bm{x}_p$ against $\bm{z}_{r, -p}$. To add the smooth regression line, like before we create an auxiliary vector $\tilde{\bm{x}}_p$ and its basis expansion $\tilde{\bm{Z}}_{p}$. With the estimated weights and loadings we can compute $\bm{z}_{r, p} = \tilde{\bm{Z}}_{p}\hat{\bm{B}}_p\hat{\bm{v}}_r$. The series of points $\tilde{\bm{x}}_p$ against $\bm{z}_{r, p}$ define a smooth curve illustrating the relationship between the predictor and the partial residuals. We will illustrate these curves in Section \ref{sec:examples}. For a fitted smooth reduced rank model there are $P \times R$ of these partial response plots. 

\section{Some simulated gauges}\label{sec:gauges}

Inspired by the literature on robust statistics \citep{leyder2026independent}, we make use of two highly nonlinear transformation functions: the \texttt{biloop}$(\bm{x})$ function and the \texttt{bowl}$(\bm{x})$ function. Both transform a vector of observations $\bm{x}$ in two vectors $\bm{u}_{1}$ and $\bm{u}_{2}$ of the same length. Both the \texttt{biloop} and \texttt{bowl} function are highly nonlinear. 

For the \texttt{biloop} function
\begin{eqnarray*}
u_{i1} &=& 4 * (1 + \cos(2 * \pi * \tanh(x_i/4) + \pi)) \ \mathrm{if} \ x_i \geq 0 \\
u_{i1} &=& -4 * (1 + \cos(2 * \pi * \tanh(x_i/4) - \pi)) \ \mathrm{if} \ x_i < 0 \\
u_{i2} &=& \sin( 2 * \pi * \tanh( x_i / 4 ) )
\end{eqnarray*}
Afterwards, we scale both, $\bm{u}_1$ and $\bm{u}_2$, to have zero mean and variance 1.
For the \texttt{bowl} function, we first define $w_i = \tanh( |x_{i}|/q)$, with $q = \sqrt{\chi^2_{0.9975, 2}}$ and subsequently
\begin{eqnarray*}
u_{i1} &=& 10 w_i^6 (1 - w_i)^2; \\
u_{i2} &=& (10 w_i^2 + (1 - w_i)^2) x_i.
\end{eqnarray*}
Again, we scale both, $\bm{u}_1$ and $\bm{u}_2$, to have zero mean and variance 1 afterwards.

\begin{figure}
    \centering
    \includegraphics[width = 1.0\textwidth]{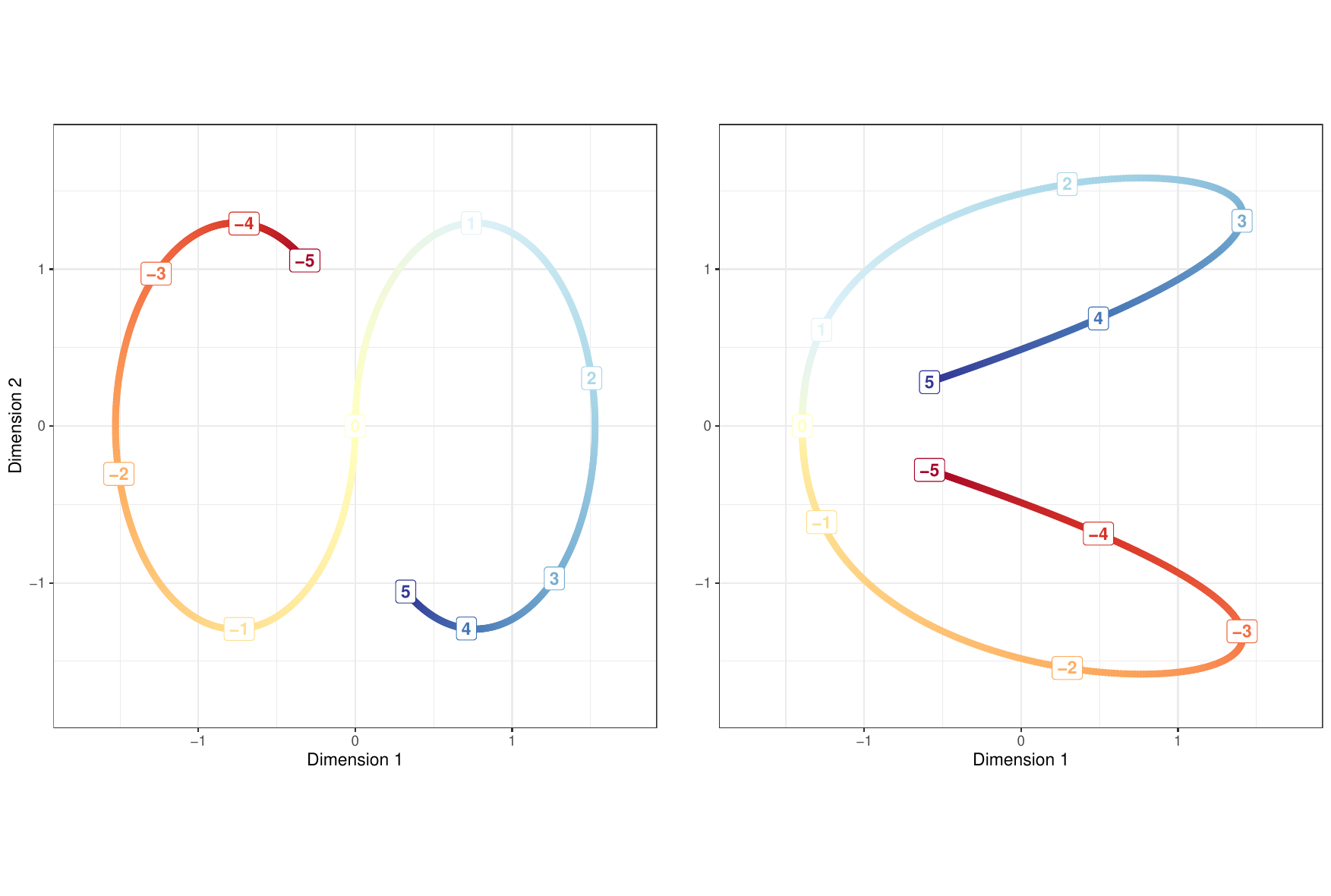}
    \caption{The two nonlinear transformations of a sequence from minus 5 to five. Left: the \texttt{biloop}-transformation; Right: The \texttt{bowl}-transformation.}
    \label{fig:transformations}
\end{figure}

We illustrate these two transformations in Figure \ref{fig:transformations}. We generated a variable running from -5 to 5 and applied the two transformations. In the left hand panel the biloop transformation is shown and in the right hand side panel the bowl  transformation. As can be verified, both transformations are highly nonlinear, and would be difficult to approximate with polynomial transformations of a variable.  

To contrast the highly nonlinear transformations, we also investigate a linear one. For the \texttt{linear} function is simply defined by
\begin{eqnarray*}
u_{i1} &=& a_1 * x_i; \\
u_{i2} &=& a_2 * x_i.
\end{eqnarray*}
for chosen values of $a_1$ and $a_2$. 

We will work with a fixed matrix $\bm{V}$ that respects the identification constraints, that is,
\[
\bm{V} = \left[
\begin{array}{rr}
0.031 & -0.470 \\
-0.970 & 0.198 \\
0.241 & 0.860 
\end{array}\right].
\]

\subsection{Experimental gauge 1}

\begin{figure}
    \centering
    \includegraphics[width = 1.0\textwidth]{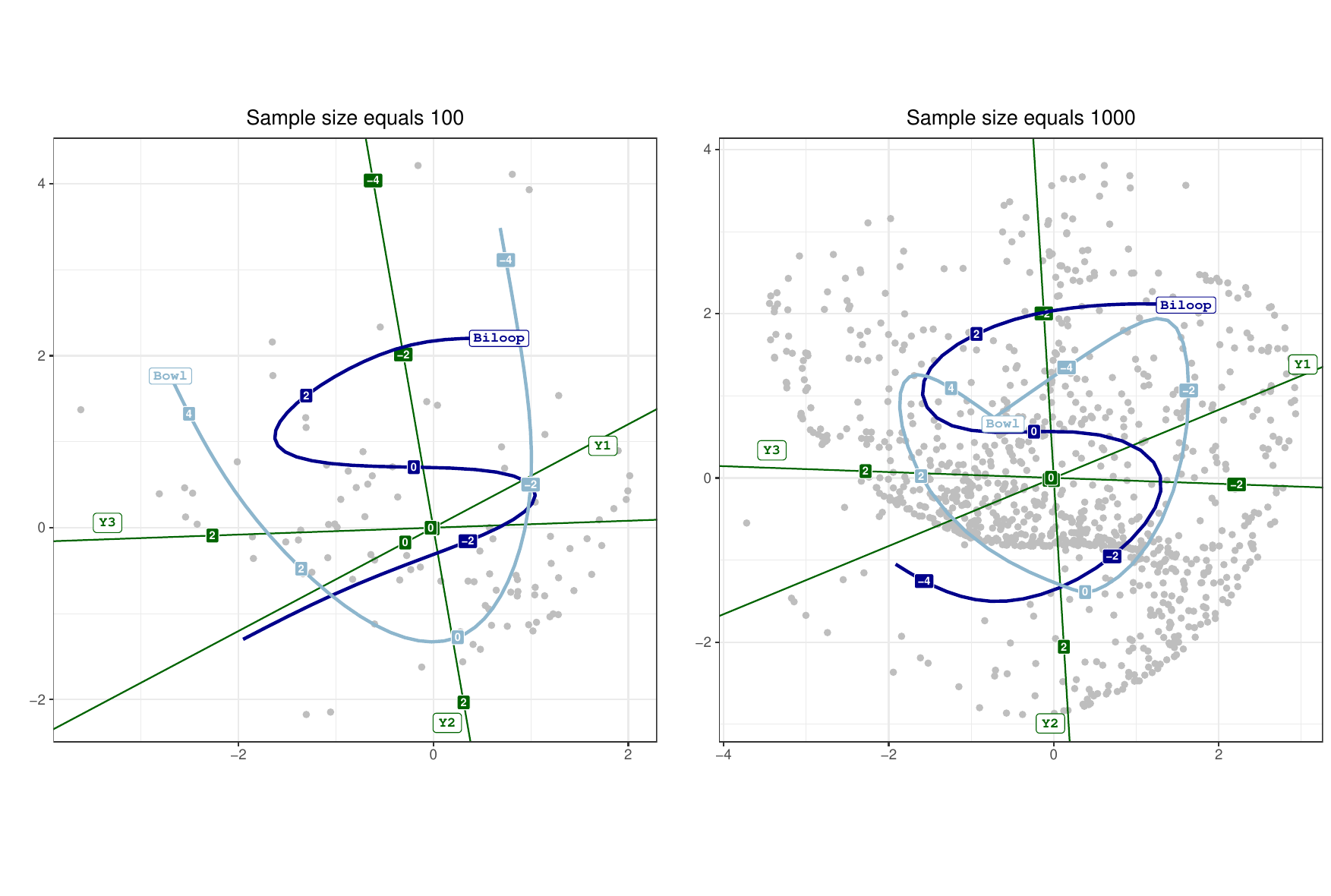}
    \caption{Result of the first experimental gauge with two highly nonlinear paths. For small sample size on the left, for large sample size on the right hand side.}
    \label{fig:gauge1}
\end{figure}

In our experiments we draw $\bm{x}_1$ and $\bm{x}_2$ from a normal distribution with mean zero and standard deviation 1.5. On the first predictor, we apply the biloop transformation and on the second the bowl transformation. To mimic the additive nature of our modeling approach the two resulting matrices with coordinates in a two-dimensional space are added, that is, we define
\[
\bm{U} = \texttt{biloop}(\bm{x}_1) + \texttt{bowl}(\bm{x}_2)
\]
and draw response variables using
\[
\bm{Y} = \bm{UV'} + \bm{E}
\]
where the elements of $\bm{E}$ are randomly drawn from an independent normal distribution, that is. $e_{ir} \sim \mathcal{N}(0, \sigma^2)$ with $\sigma^2$ chosen such that the overall explained variance is 50\%. We draw a sample of size 100 (small) and another one of size 1000 (large). 

These data with two predictor variables and three response variables are analyzed with our procedure, where we select the optimal penalty parameter using the BIC. We show the results in Figure \ref{fig:gauge1}, that consists of two displays: the results for $N = 100$ (left) and the results for $N = 1000$ (right). For the large sample size, we can see that the procedure recovers the biloop and bowl trajectories very well (upto a rotation). For the smaller sample size, the biloop is recovered well, but the bowl lacks the bending back to the origin. This is due to the small sample size because with small sample size there are just a few cases with large absolute values, that is, in our simulated data there are only three cases with smaller values than -3 and only 3 cases with values larger than 3. Therefore, there is not much support for that part of the transformation. With sample size 1000, these numbers are 25 and 26. and the recovery is much better.  

\subsection{Experimental gauge 2}

\begin{figure}
    \centering
    \includegraphics[width = 1.0\textwidth]{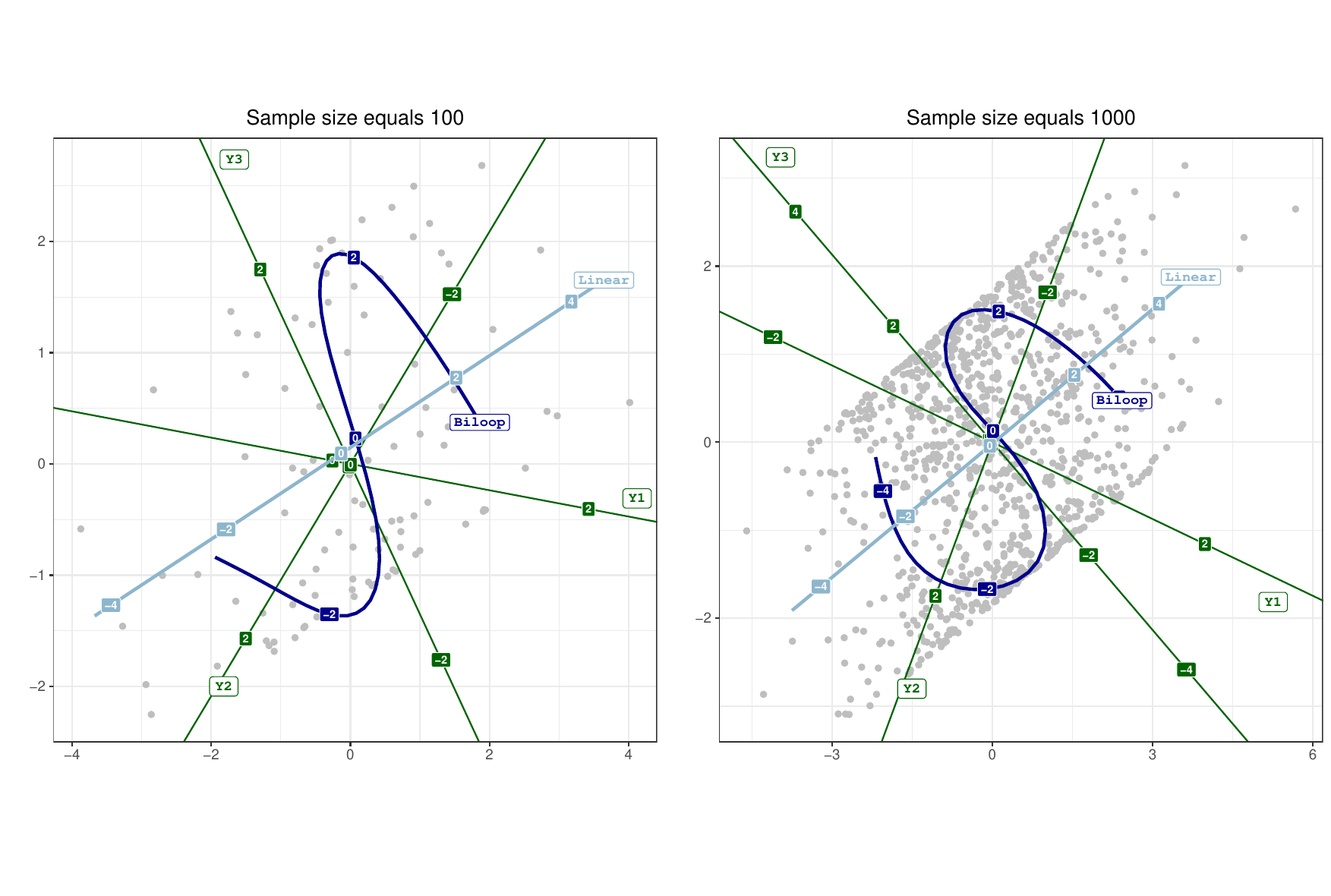}
    \caption{Result of the second experimental gauge with a highly nonlinear and a linear path. For small sample size on the left, for large sample size on the right hand side.}
    \label{fig:gauge2}
\end{figure}

In this second experiment, we draw the two vectors the same way as in the first experiment. Like before, the first predictor variable is transformed with the biloop. For the second predictor variable, we use a linear transformation with $a_1 = 0.5$ and $a_2 = -0.7$. Afterwards, we scale the resulting scores to have mean zero and variance one. The rest of the data generation is kept the same. 

The result is shown in Figure \ref{fig:gauge2}, where it can be seen that in the large sample size condition the two trajectories are recovered adequately, whereas in the small sample condition the recovery of the straight line is good but the extremes of the biloop are recovered less well. 

\subsection{Experimental gauge 3}

\begin{figure}
    \centering
    \includegraphics[width = 1.0\textwidth]{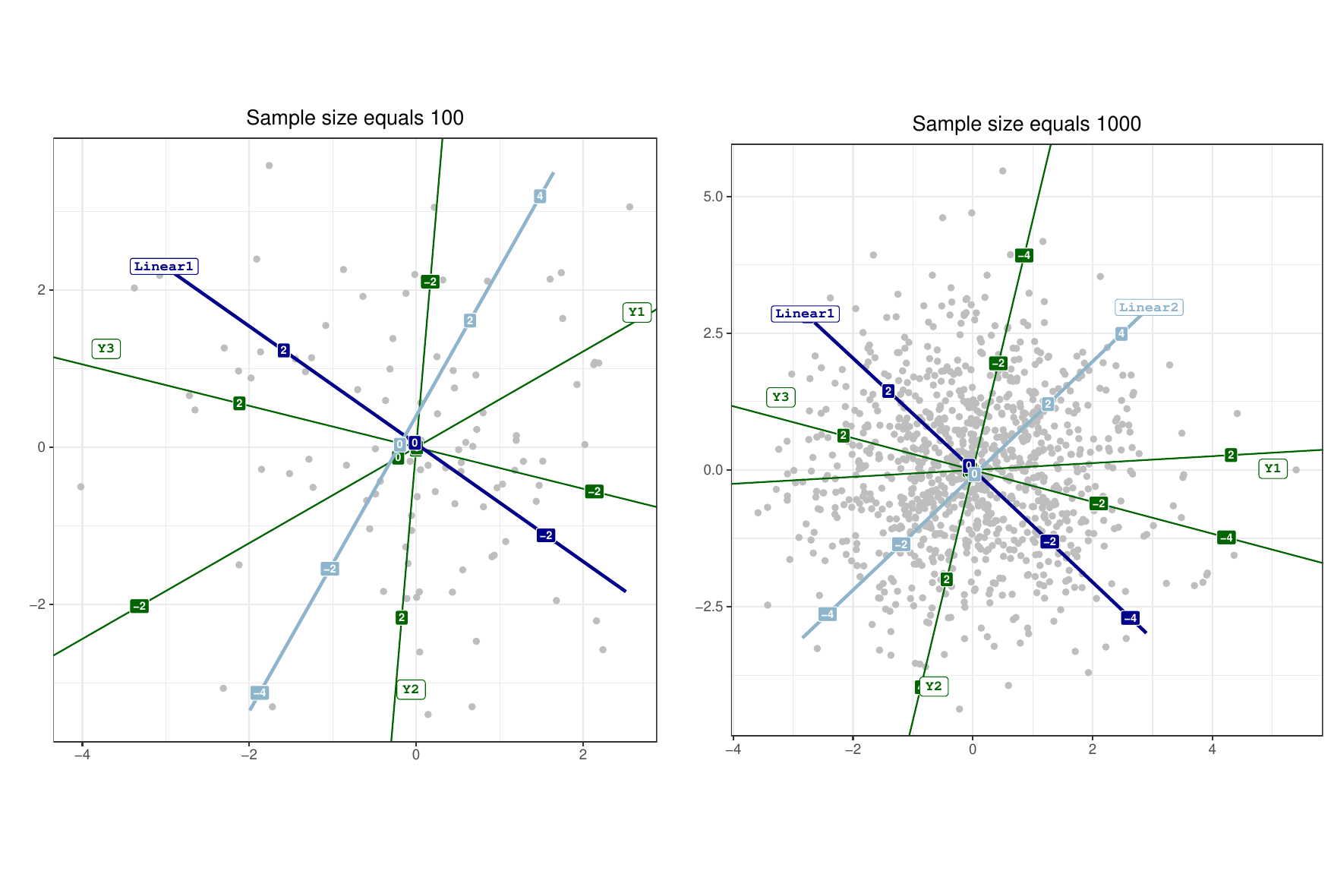}
    \caption{Result of the third experimental gauge with two linear paths. For small sample size on the left, for large sample size on the right hand side.}
    \label{fig:gauge3}
\end{figure}

In this third experiment, we draw the two vectors the same way as in the first experiment. Now, both predictor variables are linearly transformed, the first with $a_1 = 0.7$ and $a_2 = 0.4$ and the second with $a_1 = 0.5$ and $a_2 = -0.7$. Afterwards, we scale the resulting scores to have mean zero and variance one. The rest of the data generation is kept the same, again 50\% of the variance is accounted for and we draw a small and large sample. The result is shown in Figure \ref{fig:gauge3}. The result speaks for itself.  

% \subsection{Experiment 4}

% \begin{figure}
%     \centering
%     \includegraphics[width = 1.0\textwidth]{gauge4result.pdf}
%     \caption{Result of the fourth experimental gauge. For small sample size on the left, for large sample size on the right hand side.}
%     \label{fig:gauge4}
% \end{figure}

\section{Some empirical examples}\label{sec:examples}

\subsection{Example 1: Tobacco data}

The data are described in \cite{izenman2008modern} and originally from 
\cite{andersonstatistical}. The data are taken from a study on the chemical composition of
tobacco leaf samples. There are 25 observations on three response variables: rate of cigarette burn in inches per 1,000 seconds (Burnrate), percent sugar in the leaf (\%
Sugar), and percent nicotine in the leaf (\% Nicotine). We focus on two predictor variables,  
percent calcium and percent magnesium. 

We analyse the data with a rank 2 model. We contrast the model selection using the AIC and BIC. In this case, the data only has 25 observations, so we expect the penalty to have a relative large influence. Note that the log of 25 ($N$) is 3.22, so the complexity element in the BIC weights a bit more than the value 2 in the AIC.  

In this case with two predictor variables, we have a penalty parameter vector of length two. The optimal values according to the AIC are $\exp(-2)$ and $\exp(-2)$, whereas the optimal values according the BIC are $\exp(2)$ and $\exp(5)$; a large difference.

\begin{figure}
    \centering
    \includegraphics[width = 1.0\textwidth]{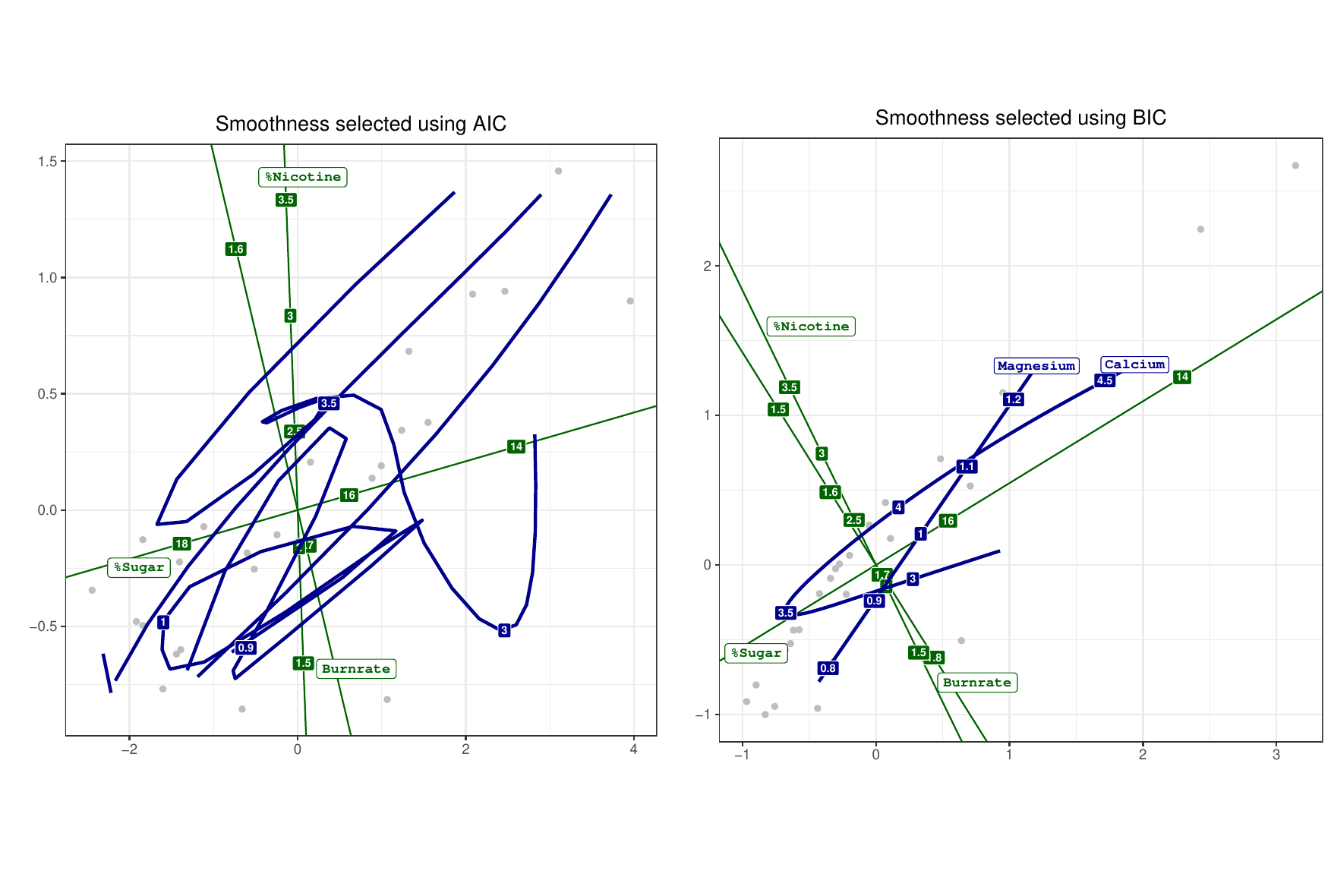}
    \caption{Biplots for Tobacco data. On the left hand side, the biplot where smoothness is determined using the AIC, whereas on the right hand side the one selected using the BIC.}
    \label{fig:tobacco}
\end{figure}

The results are shown in Figure \ref{fig:tobacco}, on the left hand side for the model selected based on the AIC and on the right had side the model selected with the BIC. The variable axes for the three response variables are very similar in the two representations. In both representations we find that the response variables \% Nicotine and Burnrate have a strong negative correlation, whereas \% Sugar is almost uncorrelated with these two.

The trajectories of the two predictor variables, however, are quite different. The trajectories in the model selected using the AIC are unsmooth, whereas the trajectories of the predictor variables in the model selected by the BIC are quite smooth. This is a result we also found in other examples, suggesting that the BIC finds a better balance between fit and smoothness. 

In the BIC biplot, we see that the the percentage of magnesium has a linear effect on the three responses; the variable axis is a straight line. The higher the percentage of magnesium the lower the percentage of sugar, the lower the burnrate, and the higher the percentage of nicotine. 

The percentage of calcium has a nonlinear effect on the response variables. With increasing percentages of calcium, the percentage of sugar goes up till the percentage of calcium is around 3.5, where-after an increase of calcium results in a decline of the percentage of sugar. The effect of percentage of calcium on the other two response variables is much lower and almost monotone. With increasing percentage of calcium the burnrate goes down and the percentage of nicotine goes up. 

\subsection{Example 2: Food and Cancer data}

This second set of data is described in \cite{takane2013constrained} and originally collected by the Food and Agriculture Organization (FAO). The observations correspond to 47 countries. The two response variables are mortality rates by lower intestine cancer (L-Intes) and rectum cancer (Rectum). We focus on three predictor variables the average daily intake of Meat, Milk, and Alcohol. The interest lies in the relationship between the cancer mortality rates and the food variables. Because the distribution of all five variables is very skew, we apply a log-transformation before the analysis. 

We fit models in dimensionality one and two, where the optimal penalty parameters are selected using the BIC. The fit statistics are shown in Table \ref{tab:food_fit}, which shows that the two dimensional model fits better, that is, the two dimensional model has lower AIC and BIC values.  

\begin{table}[!t]
\centering
\begin{tabular}{rrrrrrrr}
  \hline
 & $\log(\lambda_1)$ & $\log(\lambda_2)$ & $\log(\lambda_3)$ & NLL & ED & AIC & BIC \\ 
  \hline
  $S = 1$ & 5.00 & 5.00 & 1.75 & 112.55 & 6.79 & 238.67 & 251.23 \\ 
  $S = 2$ & 1.25 & 3.00 & 1.00 & 86.25 & 10.02 & 192.55 & 211.08 \\ 
   \hline
\end{tabular}
\caption{Fit statistics for Food and Cancer data for rank 1 and 2 models. \\ NLL = Negative log-likelihood; ED = Effective Dimension.}
\label{tab:food_fit}
\end{table}

\begin{figure}
    \centering
    \includegraphics[width = 0.9\textwidth]{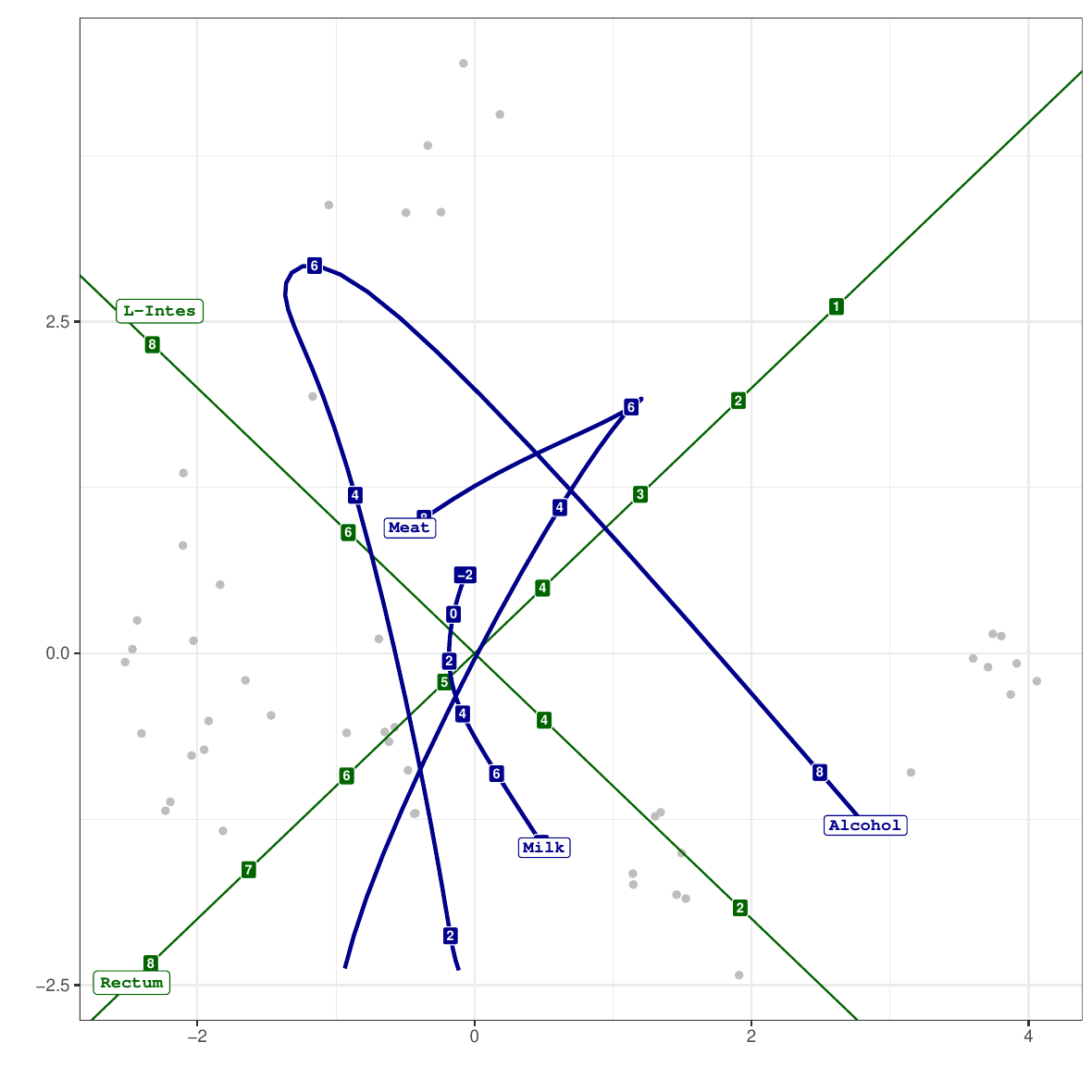}
    \caption{Biplot for Food and Cancer data.}
    \label{fig:food1}
\end{figure}

The two-dimensional model is graphically represented in a biplot in Figure \ref{fig:food1}. The two response variables lower intestine cancer and rectum cancer are represented by the straight green lines. The three predictor variables are represented by the smooth blue curves. The countries are represented by the gray dots. Unfortunately, we do not have the labels of the countries and therefore are unable to make substantive interpretations about which countries have high cancer mortality rates. 

For interpreting the effects of the food variables on the mortality rates we focus on conditional effects. Therefore, we look at the trajectories of the predictor variables  

The curve for average intake of milk is very smooth and close to a straight line. Keeping the values for intake of meat and alcohol at fixed values, we can conclude that with increasing values of intake of milk the mortality rate for lower intestine cancer goes down and the mortality rate for rectum cancer goes slightly up. To verify this conclusion, project the value marked points on the curve for intake of milk (i.e., -2, 0, 2, 4, and 6) onto the variable axes of the two response variables. 

Similarly, keeping intake of milk and meat at fixed values, we see that with increasing intake of alcohol the mortality rate for lower intestine cancer first goes up (till the value of 6) and then declines. The mortality rate for rectum cancer goes down till the average intake of alcohol equals 6 and then remains constant (i.e., the part of the curve for intake of alcohol between 6 and 8 is almost orthogonal to the variable axis for Rectum).

Finally, the conditional effect of intake of meat seems to be quite influential on the mortality rate of rectum cancer and less on lower intestine cancer. The effect of intake of meat on rectum cancer seems protective till the value of 6, with more intake of meat the chances of rectum cancer become higher. The effect of intake of meat on lower intestine cancer is monotone, higher intake leads to higher mortality rates. 

\begin{figure}
    \centering
    \includegraphics[width = 0.9\textwidth]{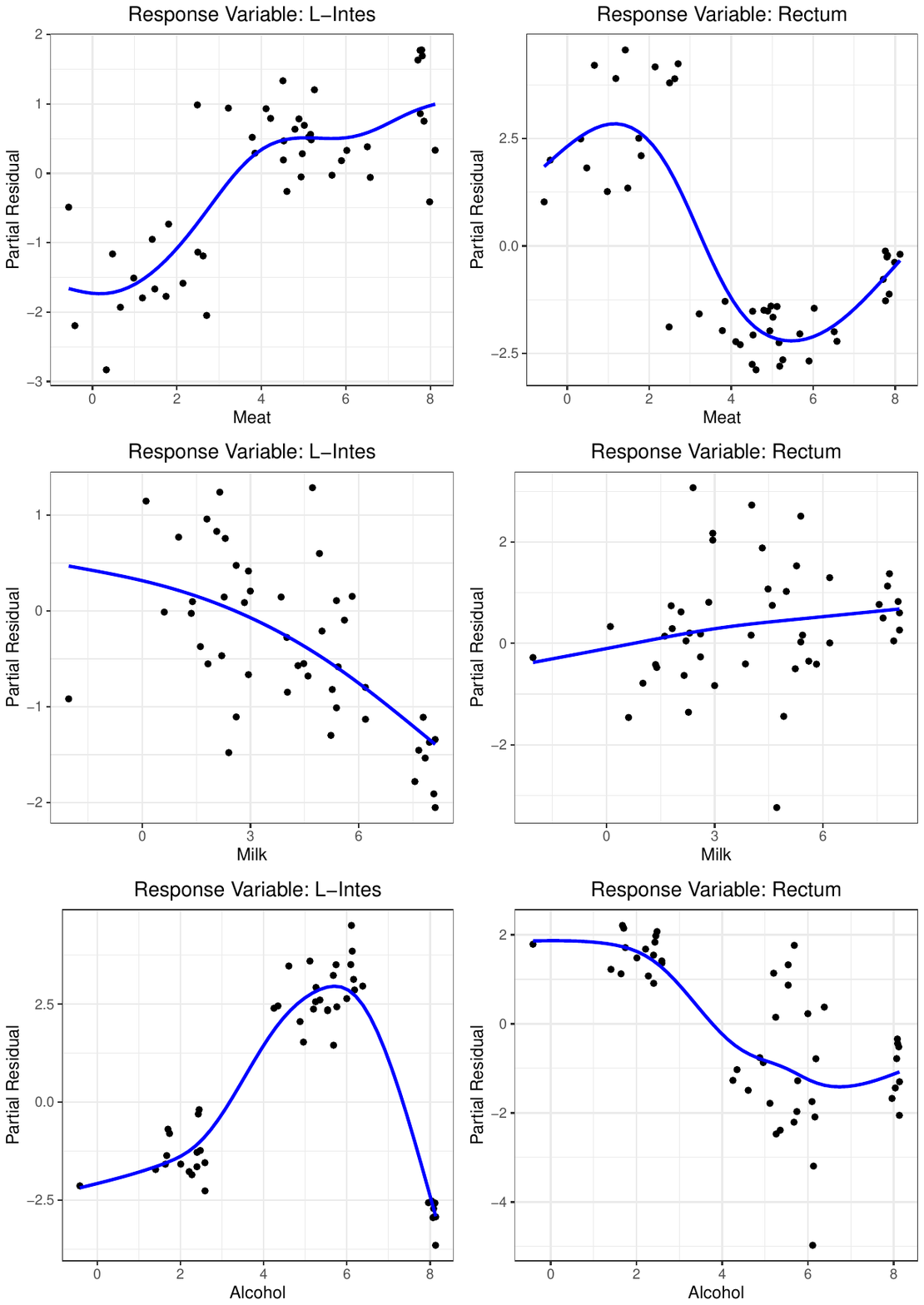}
    \caption{Partial dependence plots for Food and Cancer data. On the left hand side, the partial effects of the three predictors on lower intestine cancer. On the right hand side, the partial effects of the three predictors on the mortality rate by rectum cancer. }
    \label{fig:food2}
\end{figure}

The corresponding partial dependence plot, that are plots per predictor and response variable combination are shown in Figure \ref{fig:food2}. The upper row shows the partial dependence plots for the predictor average intake of meat, the middle row for average intake of milk, and the lower row for avarega intake of alcohol. On the left hand side the effects on mortality rates by lower intestine cancer (L-Intes) are shown, whereas on the right hand side the effects on mortality rates by rectum cancer (Rectum). We see highly nonlinear dependency patterns for meat and alcohol. The partial effect of alcohol on lower intestine cancer, for example, shows a single-peaked curve, first going up and then declining, similar to what we concluded from the biplot display. The partial effect of intake of meat on rectum cancer follows a smooth wave pattern that first slightly goes up, then downwards and at the end upwards. The effects of intake of milk on both response variables is almost linear (middle row). Note that the sharp breakpoint we found in the biplot for intake of meat is not visible in these partial dependence plots. So, what seems to be a not so smooth trajectory in the biplot transfers to a rather smooth trajectory in the partial dependence plots.

\subsection{Example 3: Ecological-Momentary-Assessment Data}

% Fried E. I., Proppert R. K. K., Rieble C. L. (2023). Building an early warning system for depression: Rationale, objectives, and methods of the WARN-D study. Clinical Psychology in Europe, 5(3), 1–25. https://doi.org/10.32872/cpe.10075

% Siepe BS, Rieble CL, Tutunji R, et al. Understanding Ecological-Momentary-Assessment Data: A Tutorial on Exploring Item Performance in Ecological-Momentary-Assessment Data. Advances in Methods and Practices in Psychological Science. 2025;8(1). doi:10.1177/25152459241286877

In this third example, we use data from \cite{fried2023} and \cite{Siepe2025}, where for a period of 85 days participants are asked every night to respond on a 7-point scale from \emph{Not at all} to \emph{Very much} to the following 10 statements:
\begin{enumerate}
\item Overall, I'm content with how my day went.
\item I was able to handle today's challenges well.
\item I am looking forward to tomorrow.
\item Today, I was able to concentrate and focus well.
\item Today, I felt connected to other people.
\item Today, I felt down or depressed.
\item Today, I had little interest or pleasure in doing things.
\item Today, I experienced physical discomfort/pain.
\item Today, it was difficult to cope with my emotions.
\item Today, I felt productive/useful. 
\end{enumerate}
Note that variables 1, 2, 3, 4, 5, and 10 are positively worded, whereas variables 6, 7, 8, and 9 are worded negatively. There is only one predictor in our analysis, that is, Day, for which we try to find a smooth trajectory and the outcome variables are the 10 variables listed above. 

% colnames(data4) = c("ID", "Day",
%                       "Concentration", "Outlook", "Discomfort", "Content",
%                       "Emo_reg", "Useful", "Anhedonia", "Connected" , "Depressed", "Coping")

We will investigate the data of 2 participants. The first participant, A, has observations on 63 days. The second participant, B, has observations on 49 days. So, they did not respond on every day. 
We allow each participant to have its own trajectory. That means that in designing our matrix $\bm{Z}$ we need to allow for an interaction between participant and the spline basis for time. This is accomplished by defining a block-diagonal structure for $\bm{Z}$. For two participants this matrix would have the following form 
\[
\bm{Z} = \left[
\begin{array}{cc}
\bm{Z}_{1} & \bm{0} \\
\bm{0} & \bm{Z}_{2} 
\end{array} \right],
\]
where $\bm{Z}_{p}$ is the B-spline basis for time (i.e., Day) for participant $p$. The matrix $\bm{Y}$ vertically concatenates the response matrices of the 2 participants. In this way, we obtain a common dependency structure of the response variables for the participants in the data set. 

\begin{figure}
    \centering
    \includegraphics[width = 1.0\textwidth]{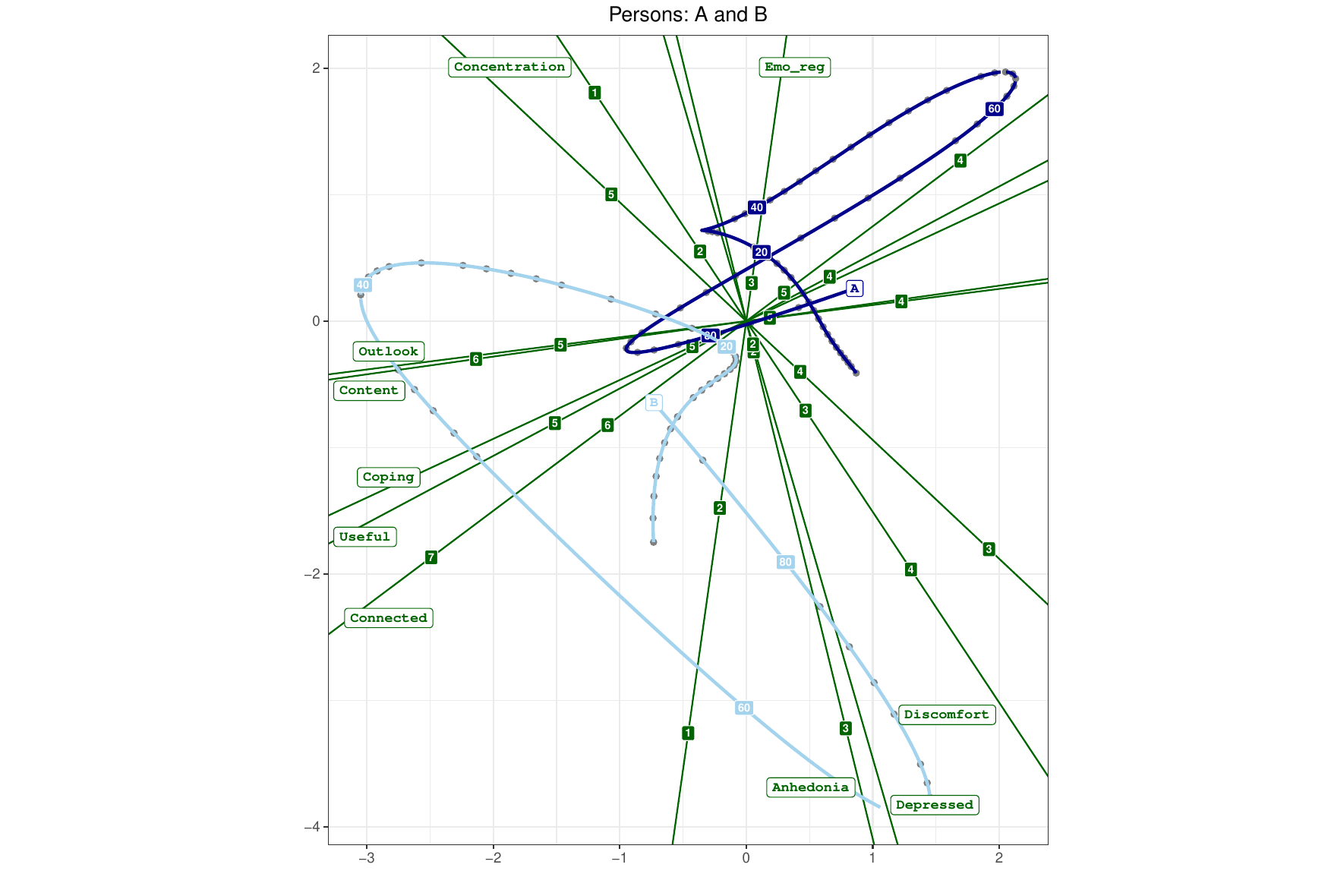}
    \caption{Biplot for ecological momentary assessment data with trajectories for two persons (A (darkblue) and B (lightblue)) in the two dimensional space representing ten response variables (darkgreen). The dots on the two smooth trajectories indicate the measurement days. The response variables are on a 1-7 scale, the numbers on the variable axes reflect these outcomes. }
    \label{fig:ema1}
\end{figure}

The result is shown in Figure \ref{fig:ema1}. Let us first consider the response variables which are represented by the straight variable axes. The variable axes cross in the origin of the plot at the average values of the variables. Each variable has a direction and a set of value indicated markers. All ten variables are observed in the range 1 to 7. The predicted values for the Emotion regulation variable are low, that is, the variable axes has only markers corresponding to the values 1, 2, and 3. In contrast, response variable Connected has markers with values 4, 5, 6, and 7, indicating that generally the two participants score high on this variable. 

We see that the response variables Connected, Useful, Coping, Content, and Outlook all have similar directions. The trajectories for these five variables is similar, that is, over the days the values for all these five variables simultaneously increase or decrease for both participants. We call these five variables Set 1. Also the variable axes for Anhedonia, Depressed, Discomfort and in opposite direction Concentration and Emotion regulation have the same directions. So, when the predicted value of Anhedonia, Depressed, and Discomfort go up, the values for the latter two, Concentration and Emotion regulation, go dow and vice versa. These latter five variables are called Set 2.

Let us check the estimated trajectories of the two participants in more detail. The trajectory of participant A lies in the higher-right quadrant, whereas the trajectory for participant B lies more in the lower left quadrant. The variables axes for Connected, Useful, Coping, Content, and Outlook, are oriented from the upper right to lower left. Therefore, participant A generally scores below the average on these variables, while participant B scores above the average. 

Participant A first remains constant on Set 1 till day 30, then the predicted values go down till day 60 where-after the values go up till around day 75 and consequently down again. For the variables in Set 2 the values go down till day 30 for Anhedonia, Depressed, Discomfort and up for Concentration and Emotion regulation, there-after the values go slowly in the other direction. 

Participant B has a different trajectory. For the variables in Set 1 the values go down till day 20, then move up till day 40 after which the values go down slowly.  For the variables in Set 2, the values go down for Anhedonia, Depressed, Discomfort and up for Concentration and Emotion regulation till around day 40, from day 40 till day 70 they go in the other direction, and after day 70 again in the original direction. As mentioned before, this participant responded on fewer days. This is apparent from the trajectory as there are few markerpoints from 40 till 60. Indeed, this participant 
responded on days 41, 46, 47, 48, 49, 50, and then 71 , so that most of the trajectory between days 50 and 70 is interpolated from the other observations. 

\begin{figure}
    \centering
    \includegraphics[width = 0.9\textwidth]{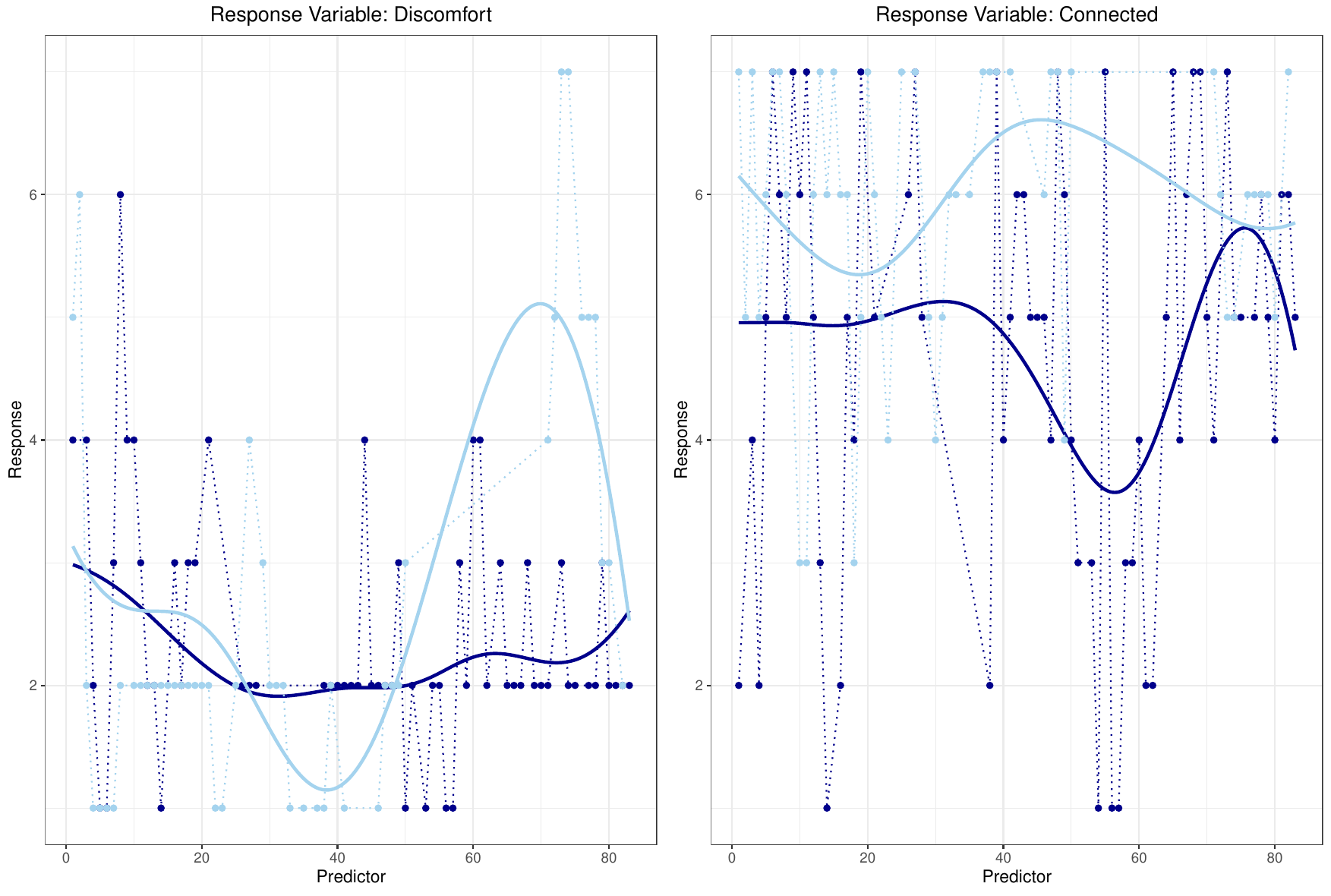}
    \caption{Predictor-response representation for response variables Discomfort and Connected with smooth trajectories for two participants (A (darkblue) and B (lightblue)). The horizontal axes represents the predictor day. Dotted lines present observed trajectories, solid lines estimated trajectories.}
    \label{fig:ema2}
\end{figure}

A detailed look at two response variables is given in the partial dependency plots in Figure \ref{fig:ema2} where on the left hand side we display the relationship between the predictor (Day) and Discomfort, while on the right hand side the relationship between Day and Connected. The points connected by the dotted lines represent the observed values. The smooth solid lines represent the trajectories. From these plots, we can derive the same conclusion about the trajectories of participant A and B as from the biplot. As these are more familiar plots their interpretation is more straightforward. With 10 response variables, we can make 10 of these plots while the triplot is a joint display of all 10 variables. These marginal plots, however, also show the observed data and as such we can see that the fit is adequate. 

\section{Conclusion and Discussion}

In this paper, we proposed to extend reduced rank regression with P-splines to relax the assumption of linear relationships between the predictors and the responses. We developed an algorithm, showed how to tune the penalty parameter with Information Criteria (AIC/BIC). 
Furthermore, we discussed triplots and partial-dependence plots for the interpretation of the resulting models. The triplots provide a global model interpretation, whereas the partial-dependence plots are specific for a one-to-one predictor response relationship. 

We then showed three experimental gauges, i.e., simulated data sets with particular characteristics to verify whether the procedure works accurately. The first gauge had to extremely non-linear trajectories for the two predictor variables, the second gauge one extremely non-linear trajectory and the other linear, and the third one  two linear trajectories. The P-spline reduced rank regression accurately retrieved the simulated functions both in a small sample size scenario and in a large sample scenario. Finally, we also showed three applications to empirical data, highlighting the value of the approach in different scenarios. With the first data set we showed the results of the model selection using the AIC and the BIC. The BIC solution is smooth whereas the AIC solution is rather irregular. This mirrors our experience with the two information criteria in applications and simulated data sets, that is, the BIC provides rather good results whereas the AIC does not regularize enough. Therefore, our preferred approach for selection of the penalty parameter is by employing the BIC.  Further research is needed on this topic. 

With the second data set, we briefly showed how rank selection and model selection can be performed. By fitting models in multiple ranks and finding for each rank the optimal set of values for the penalty parameters with resulting information criterion, we can also compare the obtained AIC or BIC for the different ranks and select the one with the lowest value. With this example, we also showed both the triplots and the partial dependence plots. 

In the third example, the data reflect time series on a set of response variables for multiple persons. As we wanted the trajectories of the different persons to be different we included an interaction between person and time in the design. We showed how to define the design matrix $\bm{Z}$ for this case, with a block-diagonal structure. This results in different trajectories of time for the different persons. Such a procedure can be used more generally, for example when the interest lies in differential development of two groups, say a treatment and a control group. Also in such a research setting, the design matrix needs to become a block-diagonal matrix with one block for group 1 (i.e., controls) and the other block for group 2 (i.e., the treated). 

For a given value of the penalty parameter the proposed algorithm is very fast and converges in a few steps. The computational complexity of the approach results from the fact that for each predictor variable we have a penalty parameter. Consider that the length of the vector of possible values for the penalty parameter is $c$, then for 2 predictors we need to run the algorithm $c^2$ times, for three predictors $c^3$ times, etc. Even with a small $c$, say 10, with three predictors we already need to cycle through 1000 runs of our algorithm. Therefore, we focused on information criteria instead of resampling methods like cross-validation for selection of the optimal penalty parameter, because the latter increases the computational burden even further. 

The AIC and BIC we used are \emph{approximate} information criteria because we made an approximation of the effective model dimension. For linear reduced rank regressions the exact degrees of freedom are difficult to determine \citep[][]{mukherjee2015degrees}. We have the extra difficulty that the penalty needs to be taken into account and therefore we generalized the approach suggested by \cite{eilers2021practical}. Further research is needed on this topic. 

In empirical example 3, the data are ordinal variables that we treated as numeric. In doing so, we made assumptions about the distances between the seven categories of each of the 10 response variables, that is, we assumed that the distances between subsequent categories are equal. This assumption is probably not tenable and it would be better to treat these variables as ordinal. Reduced rank models have been generalized for sets of binary responses \citep{yee2003reduced, derooij2023new} and sets of ordinal responses \citep{derooijbreemerwoestenburgbusing2022}. In the approaches developed by De Rooij and colleagues they employ a MM algorithm where the negative log-likelihood is iteratively majorized by a least squares function. Therefore, it should not be too difficult to generalize the approach developed in this paper to other types of response variables (although this is always easier said than done in reality). 

We developed R-functions for data analysis. These and the scripts of our gauges and empirical examples can be found on the github page of the author. 

% \section*{Statements and declarations}
% \paragraph*{Thanks:} I am grateful to Paul Eilers for reading and commenting on a previous version of the manuscript. I am grateful to Eiko Fried for giving me access to the data in the third example. 
% \paragraph*{Declarations of interest:} None
% \paragraph*{Data availability:} The R-code and the data of the first two empirical examples and the experimental gauges is available from the \href{https://github.com/mjderooij/Smooth-Reduced-Rank-Regression}{github page} of the author: . 
% The data from the third empirical example are obtained from Prof.dr. Eiko Fried, who should be contacted for a request of the data. 
% \paragraph*{Funding:} This research received no specific grant funding form any funding agency, commercial or not-for-profit sectors.
% \paragraph*{Competing interests:} The authors has no competing interests to declare. 
% \paragraph*{Author contributions:} As this is a single-author paper, all work has been done by this author. 
\bibliographystyle{apalike}
\bibliography{melodic.bib}  
\end{document}